\documentclass[12pt]{article}
\usepackage{graphicx}
\begin{document}

\input epsf

\begin{center}
{\Large\bf 
Sort Real Numbers in $O(n\sqrt{\log n})$ Time\\ 
and Linear Space}

\medskip

{\it Yijie Han}

\medskip

School of Computing and Engineering

University of Missouri at Kansas City

Kansas City, Missouri 64110

hanyij@umkc.edu\

\end{center}

\begin{abstract}
We present an $O(n\sqrt{\log n})$ time and linear space algorithm for sorting real numbers.
This breaks the long time illusion that real numbers have to be sorted by comparison sorting and take $\Omega (n\log n)$ time to be sorted. 

\noindent
Keywords: Analysis of algorithms, sorting, comparison sorting, integer sorting, sort real numbers.
\end{abstract}

\baselineskip=6.1mm

\section{Introduction}
Sorting is a fundamental problem in computer science and is used almost everywhere in programming. Currently sorting can be classified as comparison sorting
and integer sorting. It is well known that comparison sorting has
$\theta (n \log n)$ time \cite{MIT} (logarithms in this paper have base 2). Integer sorting, on the other hand, is currently known to have time 
$O(n\log \log n)$ and linear space \cite{HanSort,Han1}. This bound is
for conservative integer sorting \cite{KR}, i.e. the word size is $\log (m+n)$
bits if we are to sort $n$ integers in $\{0, 1, ..., m\}$. Nonconservative 
integer sorting, where word size can be larger than $\Omega (\log (m+n))$ bits,
can sort integers faster. Kirkpatrick and Reisch \cite{KR} show that when
word size is $O(m+n)$ bits integers can be sorted in linear time. 
We have shown \cite{HS1,HS2} that when word size is $O(\log n\log (m+n))$ bits integers can be 
sorted in linear time.

It has been a long time illusion that real numbers cannot be sorted by integer sorting and they have to be sorted by comparison sorting. All papers known to us before this paper cite sorting real numbers with $\Omega (n\log n)$ time complexity. In particular many problems in computational geometry has upper or lower bounds
of $O(n\log n)$ time because of the lower bounds of $\Omega (n\log n)$ time of sorting
$n$ points on plane or in space. 

In 2011 we submitted a proposal titled ``Integer sorting and integer related computation'' to NSF and in this proposal we wrote ``Now is probably the right time to investigate the relation between integer based
algorithms and real-value based algorithms and to study if it is possible to convert a
real-value based algorithm to an integer based algorithm and if it is possible how to design
an algorithm to convert it. At the best possible situation we expect that such conversion will
not bring time loss and thus linear time algorithm for conversion is sought. Such research will
bring many surprising results. For example, lower bounds for many problem are derived based on the
lower bounds for comparison sorting. Thus if real-value based sorting can be converted to integer based
sorting then these lower bounds derived before cannot hold.'' \cite{Prop11}.
In 2012 we submitted a proposal titled ``Serial and Parallel Sorting Algorithms with Applications'' to NSF and in this proposal in addition we wrote ``Note that real values need not necessarily to
be sorted by comparison sorting. The $\Omega (n\log n)$ lower bound for sorting is for comparison based sorting.
It may be possible that real values can be sorted by non-comparison based sorting methods.'' \cite{Prop12}. These are the earliest records we can trace for the formation of our thoughts of sorting real numbers using a non-comparison based
sorting algorithm.

In this paper we show that
for sorting purpose, real numbers can be converted to integers in 
$O(n\sqrt{\log n})$ time and thereafter be sorted with a conservative integer sorting algorithm in $O(n\log \log n)$ time \cite{HanSort,Han1} or with a nonconservative integer sorting algorithm in $O(n)$ time \cite{HS1,HS2,KR}. This result is fundamental as it breaks the illusion that real numbers have to be sorted by comparison sorting. This result will also enable many problems depending on sorting real
numbers to
be reformulated or their complexity reevaluated. Besides, problems such as hashing for real numbers, storing real numbers, comparison for real numbers, etc., 
needs to be studied or restudied. 

We use an extended RAM \cite{AHU} 
model for our computation. The model of computation we used here is the
same model used in computational geometry. As in many cases of  
algorithms in computational geometry where assumptions are made that a variable
can hold a real value, our model of computation also assumes this. Addition, subtraction, multiplication,
division, indexing and shift take constant time. The shift operation in our
algorithm is always has the form of $1\leftarrow i$ and therefore can be
replaced by the power operation of $2^i$.  
We also assume that the
floor $\lfloor \; \rfloor$ and ceiling $\lceil \; \rceil$ for a real value can be 
computed in constant time, these comes from the cast operation (which is the
floor operation) that cast a real value to an integer. These
assumptions are assumed in the computational geometry.  

We assume that a variable $v$ holding a real value has arbitrary precision.
We assume that each variable $v$ can hold an integer of finite and arbitrary number of bits. All these assumptions are natural and assumed in computational geometry.      

We may assume that for a nonnegative integer $m$, 
$exp(m)=\min \{ 2^i | 2^i \geq m \}$ 
 can be computed in constant time. This is can be achieved as
in floating point normalization and then taking the exponent, i.e. to normalize
$1/m$ and then taking the exponent. This 
assumption is for convenience only and not a must in our algorithm.
We will call this assumption the normalization assumption. We will show how
our algorithm will work with and without normalization assumption.
 
\section{Converting Real Numbers to Integers for the Sorting Purpose}
We assume that input real numbers are all positive as we can add a number
to every one of them to make them positive. We then scale them such that every number has value in $(0, 1)$ as this can be done by divide each number
by a large number. These operations do not affect the relative order of the
numbers.  

For two real numbers $1> m_1 > m_2>0$, we need to have an integer $L(m_1, m_2)$
such that $\lfloor m_1L(m_1, m_2)\rfloor \neq \lfloor m_2L(m_1, m_2) \rfloor$.  With the normalization assumption we will let $L(m_1, m_2)=2exp(\lfloor 1/|m_1-m_2| \rfloor)$. Without the normalization 
assumption we will let $L(m_1, m_2)=2^{\lfloor 1/|m_1-m_2| \rfloor}$. We had 
attemptted to use $L(m_1, m_2)=\lceil 1/|m_1-m_2| \rceil$ but it did not work out,
as for two integers $A>A^\prime >0$ ($A$ and $A^\prime$ are obtained as $\lceil 1/|a-b| \rceil$) we may have that $\lfloor Am_1 \rfloor = \lfloor Am_2 \rfloor$ and $\lfloor A^\prime m_1 \rfloor \neq \lfloor A^\prime m_2 \rfloor$.
 
Let integer $f=2^i$ be a factor (similar to $L(m_1, m_2)$). For $m$ distinct integers and an integer $a$ in them 
represents the approximation of a real value $r(a)$ such that
$a=\lfloor r(a)f \rfloor$. 
We place these $m$ integers in an array $I$ of size $2^i$ with integer $a$
placed in $\lfloor r(a)f \rfloor$. Since $1 > r(a) >0$ and therefore 
$0 \leq \lfloor r(a)f\rfloor < 2^i$.
Then for
a real number $r_1$ we can check whether $\lfloor r_1f \rfloor$ is occupied by one
of these $m$ integers. If $\lfloor r_1f\rfloor$ is vacant then we can use 
integer $a_1=\lfloor r_1f \rfloor$ to represent $r_1=r(a_1)$ 
and now we have $m+1$
distinct integers. This can proceed until we find that $\lfloor r_1f \rfloor$ is
occupied. 

When $\lfloor r_1f \rfloor$ is occupied by integer $a$ then we compare 
$r_1$ and $r(a)$ and if they are equal then we can take $r_1$ out of our sorting
 algorithm. Thus we assume that they are not equal. We 
can then get $f_1=L(r_1, r(a))$. 
This means $\lfloor r_1f_1 \rfloor \neq \lfloor r(a)f_1 \rfloor$.
If we then represent
$r_1$ by $\lfloor r_1f_1 \rfloor$ and represent $r(a)$ by $\lfloor r(a)f_1 \rfloor$ then we
can distinguish between $r_1$ and $r(a)$ for the sorting purpose.

The problem is that now $f_1>  f$. Thus to test out next real number $r_2$
we have to test out both $\lfloor r_2f \rfloor$ and $\lfloor r_2f_1 \rfloor$. We say that we
are testing at two different levels, level $f$ and level $f_1$. As we proceed, the number of levels
will increase and we have to maintain the complexity for testing to within
$o(n\log n)$ time. The two levels we have to test now are denoted by level 
$f$ and level $f_1$. If there are $l$ levels we need to test we will have
these $l$ levels sorted and maintained in a stack $S$.

Table $I$ will splits into $l$ tables with one table $I_{l'}$ maintained for the integers at level $l'$.
If for two real numbers $r_1$ and $r_2$ we have that $\lfloor r_1l' \rfloor = \lfloor r_2l' \rfloor$ then we keep only one copy of them in $I_{l'}$. Thus for $l$ levels there are
$l$ tables. For example, if we maintain levels $0, 2^5, 2^{10}, 2^{50},
2^{100}, 2^{300}$, 
then there
are 6 tables and $S[0]=0, S[1]=2^5, S[2]=2^{10}, S[3]=2^{50}, S[4]=2^{100}, S[5]=2^{300}$. We use variable $top$ to store the index of the topmost element in $S$.     

At any moment the real numbers we have examined are inserted into $I_l$ tables
and they form a tree $T$ as shown in Fig. 1.

\begin{figure}
\epsfxsize=4in
\centerline{\epsffile{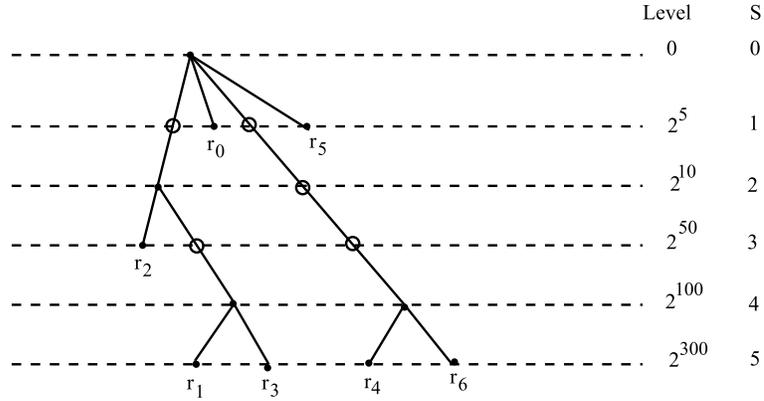}}
\caption{Real numbers are inserted according to Algorithm Insert. Circled position has no number inserted. Dotted position has integer and/or real number inserted. Real numbers are inserted at leaves.}
\end{figure}

Note that, for a real number $r$, if $l_1<l_2$ are two of the levels we maintain in $T$. Then if $I_{l_2}[\lfloor rl_2\rfloor] $ is not vacant (occupied) then
$I_{l_1}[\lfloor rl_1 \rfloor]$ must be not vacant (occupied).
We call this the transitivity property. Note that in the first version of our
algorithm presented here the transitivity proper is not kept throughout our
algorithm, but we will assume that it is kept. If the transitivity property is ketp then in Fig. 1. every circled position as well as every dotted position will have an integer inserted. We will show later how to modify our algorithm so that the transitivity
property is virtually kept. If we use virtual transitivity then in Fig. 1. only dotted positions have integer and/or real numbers inserted and circled positions have no integers inserted. To satisfy the virtual transitivity, for every node
$a$ (an internal node or a leaf, i.e. a dotted node in Fig. 1.), the following condition must be satisfied:\\

\noindent
Let $i(a)$ be the index of the level where $a$ lies in $T$, i.e., $a$ is at level $S[i(a)]$. Let $l(a)$ be any real number at a leaf of $a$ in $T$.\\
for($levelindex=0; levelindex <= \lfloor \log top \rfloor; levelindex+\! +$)\\
$\{$\\
\hspace*{0.2in}if($i(a) \; {\rm mod}\; 2^{levelindex}==0 \; \&\& \; i(a) \; {\rm mod} \;
2^{levelindex+1} \; !\! =\; 0$)\\
\hspace*{0.2in}$\{$\\
\hspace*{0.4in}$i(a)=i(a)-2^{levelindex};$\\
\hspace*{0.4in}$\lfloor l(a)S[i(a)]\rfloor$ must be a node in $T$;\\
\hspace*{0.2in}$\}$\\
$\}$\\

For two positive real numbers $0 <r_1, r_2<1$, we will say $r_1$ and $r_2$
match at level $l$ if $\lfloor r_1 l\rfloor=\lfloor r_2l \rfloor$. Let $L_S(r_1, r_2)=
\max\{ l| l \in S$ (i.e. there is an $i \leq top$ such that $l=S[i]$) and $r_1$ and $r_2$ match at level $l$ $\}$.
Let $L_{maxS}(r)=\max \{ L_S(r, a)| a $ is a previous input real number (i.e.
$a$ has already been inserted into $I_l$ tables) $\}$. The real number $a$
that achieves $L_{maxS}(r)$ is denoted by $match(r)$, i.e. $L_{maxS}(r)=L_S(r, match(r))$. 
 
For the next real number $r^\prime$ we will search on $S$ as follows:\\

\noindent
Alorithm Match($r^\prime$)\\
Input: $r^\prime$ is the next input real number to be inserted into $I_l$ tables.\\
Output: $r_0$ and $L$. $r_0$ is $match(r^\prime)$ and $L=L_S(r_0, r^\prime)$.
\\
Let $S[top]$ be the topmost element in $S$.\\
$levelindex=0;$\\
for$(i=\lfloor \log top \rfloor; \; i>=0;\; i-\! -)$  //$\lfloor \log top \rfloor$ is computed in $O(\log top)$ time.\\
$\{$\\
\hspace*{0.2in}if($levelindex+2^i <= top$ $\&\&$ $\lfloor r^\prime S[leveindex+2^i] \rfloor$ is a node in $T$\\
\hspace*{0.4in}(i.e. $I_{S[levelindex+2^i]}[\lfloor r^\prime S[levelindex+2^i] \rfloor]$ is occupied.))\\
\hspace*{0.2in}$\{$\\
\hspace*{0.4in}$levelindex=levelindex+2^i;$\\
\hspace*{0.2in}$\}$\\
\hspace*{0.2in}$i=i-1;$\\
$\}$\\
$L=S[levelindex]$, $r_0$ is a real number matched $r^\prime$ at level $S[levelindex]$;\\
 
Thus in $O(\log top)$ time we will either find a vacant position at the smallest
 level $S[levelindex+1]$ for $r^\prime$ (i.e. $I_{S[levelindex+1]}[\lfloor 
r^\prime S[levelindex+1] \rfloor ]$ is vacant; or we will find that $r^\prime$ matches a real number
at $S[top]$ and in this case we need add a new level onto $S$. 
We will insert $r^\prime$ into $I_l$ tables as follows: \\

\noindent
Algorithm Insert($r_0$, $r^\prime$)\\
Input: $r^\prime$ is the next input real number to be inserted into $I_l$ tables.
$r_0$ is $match(r^\prime)$. \\
Let $S[top]$ be the topmost element in $S$.\\
if($L_S(r_0, r^\prime)== S[top]$) push $L(r_0, r^\prime)$ onto $S$;\\
Insert $r_0$ and $r^\prime$ at level $S[S^{-1}[L_S(r_0, r^\prime)]+1]$ in $T$;\\
$levelindex=S^{-1}[L_S(r_0, r^\prime)]+1$;\\
for$(i=0; i<=\lfloor \log top \rfloor; i+\! +)$\\
$\{$\\
\hspace*{0.2in}if($levelindex \; {\rm mod} \; 2^i ==0 \; \&\& \; 
levelindex \; {\rm mod} \; 2^{i+1}\; !\! =\;0$)\\
\hspace*{0.2in}$\{$\\
\hspace*{0.4in}Insert $r_0$ and $r^\prime$ into $I_{S[levelindex-2^i]}$ if it is not inserted there before (could be there before because $r_0$ was in the $I_l$ tables), that is: insert $\lfloor r^\prime S[levelindex-2^i] \rfloor$ into table
$I_{S[levelindex-2^i]}$ if it was not there. //At most one integer is inserted.\\
\hspace*{0.4in}$levelindex=levelindex-2^i;$\\
\hspace*{0.2in}$\}$\\
$\}$\\
if($\lfloor r_0L_S(r_0, r^\prime) \rfloor$ is not in $T$)\\
  //Make the virtual transitivity structure for the internal node $\lfloor r_0L_S(r_0, r^\prime) \rfloor$\\
$\{$\\
\hspace*{0.2in}Insert $\lfloor r_0L_S(r_0, r^\prime) \rfloor$ into $T$;\\
\hspace*{0.2in}$levelindex=S^{-1}[L_S(r_0, r^\prime)]$;\\
\hspace*{0.2in}for$(i=0; i<=\lfloor \log top \rfloor; i+\! +)$\\
\hspace*{0.2in}$\{$\\
\hspace*{0.4in}if($levelindex \; {\rm mod} \; 2^i ==0 \; \&\& \; 
levelindex \; {\rm mod} \; 2^{i+1}\; !\! =\;0$)\\
\hspace*{0.4in}$\{$\\
\hspace*{0.6in}Insert $r_0$ into $I_{S[levelindex-2^i]}$ if it is not inserted there before (could be there before because $r_0$ was in the $I_l$ tables), that is: insert $\lfloor r_0 S[levelindex-2^i] \rfloor$ into table
$I_{S[levelindex-2^i]}$ if it was not there.\\
\hspace*{0.6in}$levelindex=levelindex-2^i;$\\
\hspace*{0.4in}$\}$\\
\hspace*{0.2in}$\}$\\
$\}$\\

The description of our algorithm so far will allow us to convert real numbers
to integers for sorting purpose. However, the number of levels stored in 
$S$ and $T$ could go to $O(n)$ and thus it will take $O(n\log n)$ time to convert
$n$ real numbers to integers. What we will do is to merger multiple levels into
one level and therefore eliminate many levels in $S$. 

To merge levels $S[l], S[l+1], ..., S[top]$ (we only merge the topmost levels
in $S$ to one level) into level $L=S[top]$, we will,
for any non-vacant position in $I_{l'}[a]$ for $l'=S[l], S[l+1], ...,
S[top]$,
place $r(a)$ in $I_{S[top]}$. We then
pop off $S[top], S[top-1], ..., S[l]$ off stack $S$ and
then push $L$ onto stack $S$. The value of $top$ is now equal to $l$.
Tables $I_{S[l]}, I_{S[l+1]}, ..., I_{S[top-1]}$ will be deleted.
This takes time
$O((\sum_{i=l}^{top}n_i))$, where $n_i$ is the number of
occupied positions in $I_{S[i]}$.

We will insert the $n$ input real numbers $r_0, r_1, ..., r_{n-1}$ (scaled to within $(0, 1)$) one after another into the $I$ tables. 
Let $e=2^{\sqrt{\log n}}$. After we inserted $r_0, r_1, ..., r_{e-1}$ we will merge
all levels (call these levels level $l_{0, 0}, l_{0, 1}, ..., l_{0, e-1}$)
created to the largest level (call it level $l_{1,0}$). After
$r_e, r_{e+1}, ..., r_{2e-1}$ are inserted we will merger all levels larger than
$l_{1,0}$ (call these levels $l_{0, e}, l_{0, e+1}, ..., l_{0, 2e-1}$)
 to the current largest level $l_{1,1}$. Note that some of $r_e, r_{e+1}, ...,
r_{2e-1}$ may have been inserted into level $l_{1, 0}$ and not inserted into
tables in larger levels and therefore they
will not be merged to level $l_{1, 1}$. After we inserted $r_{2e}, r_{2e+1},
..., r_{3e-1}$ we will merge all levels larger than $l_{1,1}$ to the current
largest level $l_{1,2}$. Thus after we inserted $r_{e^2-e}, r_{e^2-e+1}, ...,
r_{e^2-1}$ we will have at most $e$ levels $l_{1,0}, l_{1,1}..., l_{1, e-1}$.
At this moment we merge all levels to the largest level and call it level
$l_{2,0}$. We repeat this loop and thus after we inserted $r_{2e^2-1}$ we
can get another level $l_{2,1}$. After we inserted $r_{e^3-1}$ we can
have $e$ levels $l_{2, 0}, l_{2, 1}, ..., l_{2, e-1}$ and we will merge
all these levels to the largest level and call it level $l_{3, 0}$, 
and so on. The procedure is:\\

\vskip 0.2in

\noindent
Algorithm Merge\\
for($i_{(\log n/\log e)-1}=0; i_{(\log n/\log e)-1}<e; i_{(\log n/\log e)-1}+\! +$)\\
$\{$\\
\hspace*{0.2in}for($i_{(\log n/\log e)-2}=0; i_{(\log n/\log e)-2}<e; i_{(\log n/\log e)-2}+ \! +$)\\
\hspace*{0.2in}$\{$\\
\hspace*{0.4in}$...\; \; ...$\\
\hspace*{0.4in}$...\; \; ...$\\
\hspace*{0.4in}for($i_2=0; i_2<e; i_2+ \! +$)\\
\hspace*{0.4in}$\{$\\
\hspace*{0.6in}for($i_1=0; i_1<e; i_1+\! +$)\\
\hspace*{0.6in}$\{$\\
\hspace*{0.8in}for($i_0=0; i_0<e; i_0+\! +$)\\
\hspace*{0.8in}$\{$\\
\hspace*{1in}Insert $r_
{(\sum_{k=1}^{(\log n/\log e)-1}e^{k}i_k)+i_0}$ into $I$ tables.\\
\hspace*{0.8in}$\}$\\
\hspace*{0.8in}Merge levels $l_{0, (\sum_{k=1}^{(\log n/\log e)-1}e^{k}i_k)}$,\\
\hspace*{0.8in}$l_{0, (\sum_{k=1}^{(\log n/\log e)-1}e^{k}i_k)+1}$,\\
\hspace*{0.8in} ...,\\
\hspace*{0.8in}$l_{0, (\sum_{k=1}^{(\log n/\log e)-1}e^{k}i_k)+e-1}$\\
\hspace*{0.8in}into level $l_{1, (\sum_{k=2}^{(\log n/\log e)-1}e^{k-1}i_k)+i_1}$;\\
\hspace*{0.6in}$\}$\\
\hspace*{0.6in}Merge levels $l_{1, (\sum_{k=2}^{(\log n/\log e)-1}e^{k-1}i_k)}$,\\
\hspace*{0.6in}$l_{1, (\sum_{k=2}^{(\log n/\log e)-1}e^{k-1}i_k)+1}$,\\
\hspace*{0.6in}...,\\
\hspace*{0.6in}$l_{1, (\sum_{k=2}^{(\log n/\log e)-1}e^{k-1}i_k)+e-1}$\\
\hspace*{0.6in}into level $l_{2, (\sum_{k=3}^{(\log n/\log e)-1}e^{k-2}i_k)+i_2}$;\\
\hspace*{0.4in}$\}$\\
\hspace*{0.4in}$...\; \; ...$\\
\hspace*{0.4in}$...\; \; ...$\\
\hspace*{0.4in}Merge levels $l_{(\log n/\log e)-3, (\sum_{k=(\log n/\log e)-2}^{(\log n/\log e)-1}e^{k-(\log n/\log e)+3}i_k)}$,\\
\hspace*{0.4in}$l_{(\log n/\log e)-3, (\sum_{k=(\log n/\log e)-2}^{(\log n/\log e)-1}e^{k-(\log n/\log e)+3}i_k)+1}$,\\
\hspace*{0.4in}...,\\
\hspace*{0.4in}$l_{(\log n/\log e)-3, (\sum_{k=(\log n/\log e)-2}^{(\log n/\log e)-1}e^{k-(\log n/\log e)+3}i_k)+e-1}$\\
\hspace*{0.4in}into level $l_{(\log n/\log e)-2, (\sum_{k=(\log n/\log e)-1}^{(\log n/\log e)-1}e^{k-(\log n/\log e)+2}i_k)+i_{(\log n/\log e)-2}}$;\\
\hspace*{0.2in}$\}$\\
\hspace*{0.2in}Merge levels $l_{(\log n/\log e)-2, (\sum_{k=(\log n/\log e)-1}^{(\log n/\log e)-1}e^{k-(\log n/\log e)+2}i_k)}$,\\
\hspace*{0.2in}$l_{(\log n/\log e)-2, (\sum_{k=(\log n/\log e)-1}^{(\log n/\log e)-1}e^{k-(\log n/\log e)+2}i_k)+1}$,\\
\hspace*{0.2in}...,\\
\hspace*{0.2in}$l_{(\log n/\log e)-2, (\sum_{k=(\log n/\log e)-1}^{(\log n/\log e)-1}e^{k-(\log n/\log e)+2}i_k)+e-1}$\\
\hspace*{0.2in}into level $l_{(\log n/\log e)-1, i_{(\log n/\log e)-1}}$;\\
$\}$\\
Merge levels $l_{(\log n/\log e)-1, 0}$,
$l_{(\log n/\log e)-1, 1}$, ..., $l_{(\log n/\log e)-1, e-1}$ into one level
$l_{\log n/\log e, 0}$;\\

The loop indexed by $i_0$ takes $O(n \log top)$ time.
After we inserted $r_{ie-1}$ for $i=1, 2, ...$, 
we will merge levels $l_{0, (i-1)e}, l_{0, (i-1)e+1}, ..., l_{0, ie-1}$.
{\bf Assume (we made an assumption here)} that it takes constant time to merge each real number in
$I_{0, (i-1)e+j}$, $0\leq j<e$, to level $l_{0, ie-1}$. 
Thus the time for the loop indexed by $i_1$ excluding the time for the loop indexed by $i_0$ is $O(n)$.
After
we inserted $r_{ie^2-1}$ for $i=1, 2, ...$, we will
merge levels $l_{1, (i-1)e}, l_{1, (i-1)e+1}, ..., l_{1, ie-1}$ which
takes $O(e^2)$ time (by our assumption). Thus the time for the loop indexed
by $i_2$ excluding the time for the loops indexed by $i_1$ and $i_0$ is $O(n)$.
In general, after we inserted 
$r_{ie^j-1}$ for $i=1, 2, ...$ and $j=1, 2, ...$, we will
merge levels $l_{j-1, (i-1)e}, l_{j-1, (i-1)e+1}, ..., r_{j-1, ie-1}$ in
time $O(e^j)$ (by our assumption). Thus the time for the loop indexed by
$i_j$ excluding the time for the loops indexed by $i_0, i_1, i_2, ..., i_{j-1}$
is $O(n)$. The last line of the algorithm that is outside
all loops takes time $O(n)$ (by our assumption). 

There are $(\log n/\log e)=\sqrt{\log n}$ loops. 
The overall time for the algorithm 
is $O(n\log n/\log e+n \log top)$ (by our assumption). 

Because there are $\log n/\log e$ loops and each outstanding loop
has at most $e$ levels, thus at any time the number of levels maintained
in $S$ is $O(e\log n/\log e)$ and thus $\log top=O(\log e+\log \log n -\log\log e)=O(\sqrt{\log n})$. 

After we inserted all real numbers we will merge all levels in $T$ to the largest level.\\

\section{Keep the Virtual Transitivity Property and
Make Our Algorithm Run in $O(n\sqrt{\log n})$ Time}

The virtual transitivity is kept by Algorithm Insert. Fig. 1. shows the structure of the tree
$T$ when virtual transitivity is kept (with circled positions have no integers inserted).
As if a real 
number is inserted at a leaf node $f$ in $T$, at most $\log top$ ancestors
of $f$ will exist in $T$ by Algorithm Insert. This structure allows us to 
insert the next real number into $T$ in $O(\log top)$ time.

Note that if we never merge the topmost levels into the topmost level, our algorithm Match and Insert will work in $O(\log top)$ time. But if we do not merge levels, $top$ will go to $O(n)$. This will make our algorithm to run in
$O(n\log n)$ time. 

The problem that merging levels brings is shown in this example. 

Suppose $r_0$ and $r_1$ matched at level $S[2^i]$ for some integer $i$ 
and they do not match at
level $S[2^i+1]$, if next we merge levels $S[2^i-1], S[2^i], ... S[top]$
into one level $S[2^i-1]$ (it value is now equal to $S[top]$), then we need to
insert $r_0$ and $r_1$ into tables at levels $S[2^{i-1}], S[2^{i-1}+2^{i-2}],
S[2^{i-1}+2^{i-2}+2^{i-3}], ..., S[2^{i-1}+2^{i-2}+2^{i-3}+...+1]$. That is to 
say, in order to merge levels for two numbers $r_0$ and $r_1$ we need
possibly spend $O(\sqrt{\log n})$ time instead of constant time when $2^{O(\sqrt{\log n})}$ levels are maintained. 
 
Also suppose $r_0$ and $r_1$ match at level 
$S[2^i+2^{i-1}+2^{i-2}+...+2]$ and does
not match at level $S[2^i+2^{i-1}+2^{i-2}+...+2+1]$ and if we merge levels
$S[2^i+1], S[2^i+2], ..., S[top]$ into one level $S[2^i+1]$ (thus the value of
$S[2^i+1]$ will become $S[top]$)  then the insertions
of $r_0$ and $r_1$ in tables at levels $S[2^i+2^{i-1}], S[2^i+2^{i-1}+2^{i-2}],
..., S[2^i+2^{i-1}+2^{i-2}...+2], S[2^i+2^{i-1}+2^{i-2}...+2+1]$ need to be removed. This will also entail $O(\sqrt{\log n})$ time instead of constant time
when $2^{O(\sqrt{\log n})}$ levels are maintained.

The $O(n\sqrt{\log n})$ time complexity for Algorithm Merge requires that 
the operations in the above two paragraphs take constant time.
 
To overcome this problem we will maintain that each internal node of $T$ has at least $\sqrt{\log n}$ real numbers at its leaves. 
For the next input real number $r^\prime$,
we first find $r_0=match(r^\prime)$ (if $r_0$ is not unique we pick anyone of them) and $L=L_{maxS}(r^\prime)$. If $\lfloor r_0 L\rfloor$ is an internal 
node in $T$, then we insert $r^\prime$ at level $S[S^{-1}[L]+1]$.
Let $(S^{-1}[L]+1) \% 2^i=0$ and $(S^{-1}[L]+1) \% 2^{i+1} \; !\! = \; 0$, where
$\%$ is the integer modulo operation. Then the parent of $\lfloor r^\prime
S[S^{-1}[L]+1]\rfloor $ in $T$ is $\lfloor r^\prime S[S^{-1}[L]+1-2^i]\rfloor$.

If $b=\lfloor r_0 L\rfloor$ is a leaf in $T$ then if the set $A$ of real numbers 
at $b$ (i.e. the real numbers $r$'s such that $\lfloor rL \rfloor=\lfloor r_0 L\rfloor$)
satisfying  $|A| < 2\sqrt{\log n}-3$, then $r^\prime$ will 
be added to the set $A$ of real numbers at $b$. 
$r^\prime$ will not look for matches at levels larger than $L$. Thus $b$ 
keeps to be a leaf.
If $|A|=2\sqrt{\log n}-3$ then we will first add $r^\prime$ to $A$ and thus
$|A|=2\sqrt{\log n}-2$. Then 
the median $m_1$ of $A$ ($m_1$ has rank $\sqrt{\log n}-1$ in $A$) is found.
Let $M$ be the multiset of real numbers in $A$ that are equal to $m_1$ (It
is a multiset because previously when we add a real number $r$ to $A$
we did not look for real numbers in $A$ that are equal to $r$.)
We get $B=(A-M)\cup \{m_1\}$.
Then $m_2$ which is the smallest real number in $B$ that is larger than $m_1$
is found.
If $L_S(m_1, m_2) ==  S[top]$ then we will do $top=top+1; S[top]=L(m_1, m_2)$. Then we will do:\\

\noindent
Algorithm Branch($m_1, m_2, l, B$)  // $S[l]=L$\\
$m_1$ and $m_2$ are the two real numbers mentioned above 
at leaf node $\lfloor m_1S[l]\rfloor$.\\
$B=(A-M)\cup\{m_1\}$ as mentioned above.\\
$levelindex=0;$\\
$index=0;$\\
for($i=\sqrt{\log n}; i>=0; i-\! -$)\\
$\{$\\
\hspace*{0.2in}if($levelindex+2^i<=S^{-1}[L_S(m_1, m_2)]+1$)  //$L_S(m_1, m_2)$ is computed in $O(\sqrt{\log n})$ time. \\
\hspace*{0.2in}$\{$\\
\hspace*{0.4in}$C[index]=levelindex+2^i$;\\
\hspace*{0.4in}$index+\! +;$\\
\hspace*{0.4in}$levelindex=levelindex+2^i;$\\
\hspace*{0.2in}$\}$\\
$\}$\\
$index-\! -;$\\
foreach($a \in B$)\\
$\{$\\
\hspace*{0.2in}$M(a)=\lfloor a L(m_1, m_2)\rfloor$;\\
$\}$\\
($SortedI[0..|B|-1], SortedR[0..|B|-1]$)=sort($M(.), B(.)$); 
//Sort real numbers in $B$ by their $M(.)$ values. This is integer sorting and takes linear time \cite{HS1}.\\
$first=0;$\\
$last=|B|-1;$\\
$index1=0$;\\
$count=|A|;$\\
$flaglastindex=false;$\\
while($SortedR[first] < m_1$ $||$ $SortedR[last]>m_1$)\\
//Branch out and make sure that internal node of $T$ has at least $\sqrt{\log n}$ real numbers at its leaves.\\
$\{$\\
\hspace*{0.2in}Add $\lfloor m_1S[C[index1]] \rfloor$ into $T$ if it is not alreay there.\\
\hspace*{0.2in}while($SortedR[first] < m_1$ $\&\&$ $\lfloor SortedR[first]S[C[index1]]\rfloor \; !\! = \; \lfloor m_1S[C[index1]] \rfloor$)\\
\hspace*{0.2in}$\{$\\
\hspace*{0.4in}$count-\! -;$\\
\hspace*{0.4in}Add $\lfloor SortedR[first]S[C[index1]]\rfloor$ into $T$ if it is not already there.\\
\hspace*{0.4in}if($count < \sqrt{\log n}$ $\&\&$ $flaglastindex == false$) $flaglastindex=true;$\\
\hspace*{0.4in}$first+\! +;$\\
\hspace*{0.2in}$\}$\\
\hspace*{0.2in}while($SortedR[last] > m_1$ $\&\&$ $\lfloor SortedR[last]S[C[index1]]\rfloor \; !\! =\; \lfloor m_1S[C[index1]] \rfloor$)\\
\hspace*{0.2in}$\{$\\
\hspace*{0.4in}$count-\! -;$\\
\hspace*{0.4in}Add $\lfloor SortedR[last]S[C[index1]]\rfloor$ into $T$ if it is not already there.\\
\hspace*{0.4in}if($count < \sqrt{\log n}$ $\&\&$ $flaglastindex == false$) $flaglastindex=true;$\\
\hspace*{0.4in}$last-\! -;$\\
\hspace*{0.2in}$\}$\\
\hspace*{0.2in}if($flaglastindex == false$) $index1+\! +;$;\\
$\}$\\

Algorithm Branch is used to branch out from a leaf node $f$ when there are $2\sqrt{\log n}-2$ real numbers at $f$. After running Algorithm Branch, each leaf
node in $T$ will have less than $2\sqrt{\log n}-2$ real numbers and each internal node will have at least $\sqrt{\log n}$ real numbers at its leaves. Note that 
Algorithm Branch converts a leaf node having $2\sqrt{\log n}-2$ real numbers to leaf nodes with less than $\sqrt{\log n}$ real numbers. Thus Algorithm Branch takes linear time, i.e. $O(n)$ if we do not merging levels.

Because each internal node of $T$ has at least $\sqrt{\log n}$ real numbers at its leaves and
therefore the two problems with merging levels we posted at the beginning of
this Section can be readily solved in linear time as we have to make changes to 
no more than $O(\sqrt{\log n})$ internal nodes in $T$ and $S$ in the operations
associated with the two problems we posted. Because each internal node in $T$ has at least
$\sqrt{\log n}$ real numbers at its leaves and thus the time for adjusting the $O(\sqrt{\log n})$ internal nodes in $T$ is made to be linear time.

As we explained in the Section 2 that the overall time for merging levels is $O(n\sqrt{\log n})$. Because there are $2^{O(\sqrt{\log n})}$ levels in $T$ (or that many elements in $S$) and therefore for the next real number $r$
to find $match(r)$ takes $O(\sqrt{\log n})$ time. 

Note that after we merged all levels to the largest level each leaf node of $T$
can have up to $2\sqrt{\log n}-3$ real numbers. The real numbers within each
leaf node of $T$ needs to be sorted to determine the largest level to which to
merge all levels into (i.e. all real numbers can be converted to different integers at this level). This can be done with comparison sorting
in $O(n\log \log n)$ time.
  
\noindent
{\bf Theorem 1:} For sorting purpose $n$ real numbers can be converted to 
$n$ integers in $O(n\sqrt{\log n})$ time. $\framebox{}$

\noindent
{\bf Corollary:} $n$ real numbers can be sorted in $O(n\sqrt{\log n})$ time. 

\noindent
{\bf Proof:} First convert these real numbers to integers in $O(n\sqrt{\log n})$ time, then sort these integers with a conservative integer sorting algorithm
in $O(n\log \log n)$ time \cite{HanSort,Han1} or with a nonconservative 
integer sorting algorithm 
in $O(n)$ time \cite{HS1,HS2,KR}. $\framebox{}$ 

\section{Sorting in Linear Space}
The algorithm we presented in previous section uses nonlinear space. To make our
algorithm to run in linear space we use the results in \cite{Andersson,AnderssonThorup,Thorup} to make our algorithm run in linear space:\\

\noindent
1.  P\v{a}tra\c{s}cu and Thorup's result \cite{Thorup}. This result allow
insertion and membership lookup in an ordered set of $n$ integers
to be performed in
$O(\log n/\log w)$ time and linear space, where $w$ is the word length (i.e. 
the number of bits
in an integer). This says that search and insert an integer into an ordered list
of integers can be done in constant time and linear space 
if $w=n^{1+\epsilon}$. 
Thus if we enforce that the bits we extracted from a real
number is greater than $n^\epsilon$, i.e. if $exp(a)<n$ then we let $b=n$ 
and if $exp(a)>n$ then we let $b=exp(a)$ then we can run our algorithm
in linear space. The problem of this approach is that this result \cite{Thorup}
requires that the floating point number normalization be done in constant
time. That is it needs the $exp()$ function to be computed in constant time.
Thus if we use \cite{Thorup} then we cannot avoid the $exp()$ operation.\\

\noindent
{\bf Theorem 2:} If each real number can use at least $w>n^{1+\epsilon}$ bits
and floating point
normalization can be done in constant time then our algorithm can sort
real numbers in $O(n\sqrt{\log n})$ time and linear space.

\noindent
{\bf Proof:} In each level in our algorithm we used indexing and nonlinear
space to find whether
any integer in this level is equal to the integer to be inserted. Now we
can use \cite{Thorup} to do this in linear space and constant time provided
$w > n^{1+\epsilon}$ and floating point normalization can be done in constant 
time.  $\framebox{}$\\

\noindent
2. Andersson's result \cite{Andersson}. This result allows insertion and membership lookup in an ordered set of $n$ integers to be performed in
$O(\log n/\log w+\log \log n)$ time and linear space. This result does not
require the $exp()$ operation to be done in constant time. Thus if we enforce
that $w > n^\epsilon$ then the insertion and membership lookup can be done in $O(\log \log n)$ time and linear space and therefore our algorithm can run in $O(n\sqrt{\log n}\log \log n)$ time and linear space. 1. and 2. require that $w> n^\epsilon$ and this can be viewed as a weakness of
these methods.

The usage of \cite{Andersson} in our algorithm is the same as in Theorem 2 except we do not need the assumption that floating point number need to be normalized
in constant time.\\ 

\noindent
{\bf Theorem 3:} If each real number can use at least $w > n^{1+\epsilon}$ bits
then our algorithm
can sort real numbers in $O(n\sqrt{\log n}\log\log n)$ time. $\framebox{}$\\

\noindent
3. Andersson and Thorup's result \cite{AnderssonThorup}. This result
allows insertion and membership lookup in an ordered set of $n$ integers
to be performed in $O(\sqrt{\log n/\log \log n})$ time and linear space. 
This result does
not require the $exp()$ operation to be performed in constant time and it
does not require that $w>n^\epsilon$. This result will make our algorithm
run in $O(n\log n/\sqrt{\log \log n})$ time and linear space. 

The usage of \cite{AnderssonThorup} in our algorithm is the same as in Theorem 2 except we do not need the assumption that $w>n^{1+\epsilon}$ and floating point number can be normalized
in constant time.\\

\noindent
{\bf Theorem 4:} Real numbers can be sorted in $O(n\log n/\sqrt{\log \log n})$ time and linear space. $\framebox{}$\\

Note that when we apply 2. or 3. here we can use conservative integer sorting
(using word of $O(\log(m+n)$ bits to sort $n$ integers in $\{ 0, 1, ..., m\}$)
with time $O(n\log \log n)$ and linear space \cite{HanSort} to replace the nonconservative integer sorting we used before in our algorithm as we can tolerate the factor of 
$\log \log n$ in our algorithm when we apply 2. or 3..
   
\section{Conclusions}
Although we showed that real numbers need not be sorted by comparison sorting, our real number sorting algorithm is not as fast as our algorithm for integer sorting. But we opened the door for the study of sorting real numbers with a non-comparison based sorting algorithm. Further research may speed up the algorithm for sorting real numbers and/or results in new paradigms, approaches, methods of treating real numbers.

\end{document}